\documentclass[aps,amsmath]{revtex4}
\usepackage{graphicx}  
\usepackage{dcolumn}   
\usepackage{amssymb}   
\def\MET{{\mbox{$E\kern-0.57em\raise0.19ex\hbox{/}_{T}$}}}
\def\met{{\mbox{$E\kern-0.57em\raise0.19ex\hbox{/}_{T}$}}}

\def\DZ{D0 }
\def\DZero{D0 }
\def\Dzero{D0 }

\def\ifb{~fb$^{-1}$}
\def\pp{$p\bar{p}$}

\def\lmet{$WH\rightarrow \ell\kern-0.45em\raise0.19ex\hbox{/} \nu b\bar{b}$}

\def\tevE{$\sqrt{s}=1.96$~TeV}

\begin{document}

\rightline{FERMILAB-CONF-11-348-E}
\rightline{CDF Note 10510}
\rightline{\DZ Note 6203}
\vskip0.5in

\title{Combined CDF and \DZ Searches for the Standard Model Higgs Boson
Decaying to Two Photons with up to 8.2 fb$^{-1}$\\[2.5cm]}

\author{The TEVNPH Working Group\footnote{The Tevatron
New-Phenomena and Higgs Working Group can be contacted at
TEVNPHWG@fnal.gov. More information can be found at http://tevnphwg.fnal.gov/.}}

\affiliation{\vskip0.3cm for the CDF and \DZ Collaborations\\ \vskip0.2cm
\today}
\begin{abstract}
\vskip0.3in
We combine results from CDF and D0's direct searches for the standard model (SM)
Higgs boson ($H$) produced in \pp~collisions at the Fermilab Tevatron at $\sqrt{s}=1.96$~TeV, focusing
on the decay $H\rightarrow\gamma\gamma$.  We compute upper limits on the Higgs boson production
cross section times the decay branching fraction in the range $100<m_H<150$~GeV$/c^2$, and
we interpret the results in the context of the standard model.
We use the MSTW08 parton distribution
functions and the latest theoretical cross section predictions
when testing for the presence of a SM Higgs boson.  With datasets corresponding to 7.0\ifb\ (CDF)
and 8.2\ifb\ (D0), the 95\% C.L. upper limits on Higgs boson production is a factor of 10.5
times the SM cross section for a Higgs boson mass of 115~GeV/$c^2$.
\\[2cm]
{\hspace*{5.5cm}\em Preliminary Results}
\end{abstract}

\maketitle

\newpage
\section{Introduction} 

The search for a mechanism for electroweak symmetry breaking, and in
particular for a standard model (SM) Higgs boson, has been a major
goal of particle physics for many years, and is a central part of the
Fermilab Tevatron physics program, and also that of the Large Hadron Collider.
Recently, the ATLAS and CMS
Collaborations at the Large Hadron Collider have released results on searches for the
standard model Higgs boson
decaying to $W^+W^-$~\cite{atlasww,cmsww}, $ZZ$~\cite{atlaszz},
and $\gamma\gamma$~\cite{atlasgammagamma,atlasgammagammaupdate}.
Both the CDF and \Dzero Collaborations
have performed new combinations~\cite{cdfHWW,DZHiggs,prevcomb} of multiple
direct searches for the SM Higgs boson.
The sensitivities of these new combinations significantly exceed those
of previous combinations~\cite{prevhiggs,WWPRLhiggs}.

In this note, we combine the most recent results of
searches for the SM Higgs boson produced in \pp~collisions at~\tevE, where the
decay of the Higgs boson is restricted to pairs of photons: $H\rightarrow\gamma\gamma$.
We consider all four SM production modes, gluon-gluon fusion: $gg\rightarrow H$,
associated production of a Higgs boson with a $W$ boson: $q\bar{q}\rightarrow WH$,
associated production of a Higgs boson with a $Z$ boson: $q\bar{q}\rightarrow ZH$,
and vector-boson fusion: $q\bar{q}\rightarrow q^{\prime}\bar{q}^{\prime}H$.
The CDF search~\cite{cdfHgg} is performed with 7.0\ifb\ and the D0 search~\cite{dzHgg}
is performed with 8.2\ifb\ of collision data.

The data for the CDF search are divided into four categories:
Central-Central (CC), Central-Plug (CP), Central-Central Conversion (CC Conv), and Central-Plug conversion (CP Conv),
which are based on whether the photon(s) are detected in the central electromagnetic (EM)
calorimeter ($|\eta|<1.1$) or the plug EM calorimeter ($1.2<|\eta|<2.8$),
and whether or not one of the central photons
is detected via its conversion products $e^+e^-$.  The reconstructed mass resolution is
approximately 2.8~GeV/$c^2$ for the CDF channels.  The acceptances for Higgs boson decay
are approximately 13\% for the Central-Central category, 16\% for the Central-Plug category,
2.9\% for the Central-Central conversion category, and 1.8\% for the Central-Plug Conversion
category, varying by $\pm 10$\% (relative) depending on the mass of the Higgs boson and the
production mechanism.  We assume the SM mixture of the production mechanisms.  The discriminant
variable is the reconstructed candidate mass $m_{\gamma\gamma}$, and the background is parameterized
as a smooth function of $m_{\gamma\gamma}$ which is fit to the data outside of a window centered on
the signal under test.

The D0 search requires events with at least two photon candidates with 
$E_T > 25$~GeV and $|\eta|<1.1$. The diphoton mass resolution is $\sim 3$~GeV/$c^2$. The
contribution of jets misidentified as photons is reduced by combining
information sensitive to differences in the energy deposition from these
particles in the tracker, calorimeter and central preshower in a neural
network. In this latest iteration of the analysis boosted decision trees,
rather than the diphoton invariant mass, are used as the final discriminating
variable. The transverse energies of the leading two photons along with
azimuthal opening angle between them and the di-photon invariant mass and
transverse momentum are used as input variables. A sizeable improvement in
sensitivity beyond that achieved with the invariant mass is obtained.

\section{Signal Predictions and Uncertainties}
\label{sec:theory}

We normalize our Higgs boson signal predictions to the most recent highest-order calculations
available, for all production processes considered.  The largest production cross section, $\sigma(gg\rightarrow H)$,
is calculated at next-to-next-to-leading order (NNLO) in QCD with soft gluon resummation to
next-to-next-to-leading-log (NNLL) accuracy, and also includes
two-loop electroweak effects and handling of the running $b$ quark
mass~\cite{anastasiou,grazzinideflorian}.  The numerical values in Table~\ref{tab:higgsxsec} are
updates~\cite{grazziniprivate} of these predictions with $m_t$ set to 173.1~GeV/$c^2$~\cite{tevtop10}, and an exact treatment
of the massive top and bottom loop corrections up to next-to-leading-order (NLO) + next-to-leading-log (NLL) accuracy.
The factorization and renormalization scale choice for this calculation is $\mu_F=\mu_R=m_H$.
These calculations are refinements of earlier NNLO calculations of the $gg\rightarrow H$
production cross section~\cite{harlanderkilgore2002,anastasioumelnikov2002,ravindran2003}.
Electroweak corrections were computed in Refs.~\cite{actis2008,aglietti}.
Soft gluon resummation was introduced in the prediction of the
$gg\rightarrow H$ production cross section in Ref.~\cite{catani2003}.

The $gg\rightarrow H$ production cross section depends strongly on
the gluon parton density function, and the accompanying value of
$\alpha_s(q^2)$.  The cross sections used here are calculated
with the MSTW08 NNLO PDF set~\cite{mstw2008}, as recommended by the PDF4LHC working group~\cite{pdf4lhc}.
We follow the PDF4LHC working group's prescription
to evaluate the uncertainties on the $gg\rightarrow H$ production cross section
due to the PDFs.  This prescription is to evaluate the predictions of $\sigma(gg\rightarrow H)$ at NLO
using the global NLO PDF sets CTEQ6.6~\cite{cteq66}, MSTW08~\cite{mstw2008}, and NNPDF2.0~\cite{nnpdf}, and to
take the envelope of the predictions and their uncertainties due to PDF+$\alpha_s$ for the three sets as the uncertainty range
at NLO.  The ratio of the NLO uncertainty range to that of just MSTW08 is then used to scale the
NNLO MSTW08 PDF+$\alpha_s$ uncertainty, to estimate a larger uncertainty at NNLO.  This procedure roughly doubles the
PDF+$\alpha_s$ uncertainty from MSTW08 at NNLO alone.

We also include uncertainties on $\sigma(gg\rightarrow H)$  due to uncalculated higher order processes by following the
standard procedure of varying the factorization renormalization scales up and down together by a factor $\kappa=2$.
We treat the scale uncertainties
as 100\% correlated between jet categories and between CDF and D0, and
also treat the PDF+$\alpha_s$ uncertainties in the cross section as correlated
between jet categories and between CDF and D0.  We treat however the PDF$+\alpha_s$ uncertainty
as uncorrelated with the scale uncertainty.

We include all significant Higgs boson production modes in our searches.
Besides gluon-gluon fusion through virtual quark loops, we include Higgs boson production in association with a $W$
or $Z$ vector boson, and vector boson  fusion (VBF).  We use the $WH$ and $ZH$
production cross sections computed
at NNLO in Ref.~\cite{djouadibaglio}.  This calculation starts with the NLO calculation of
{\sc v2hv}~\cite{v2hv} and includes NNLO QCD contributions~\cite{vhnnloqcd}, as well
as one-loop electroweak corrections~\cite{vhewcorr}.
We use the VBF cross section computed at NNLO in QCD in Ref.~\cite{vbfnnlo}.  Electroweak corrections
to the VBF production cross section are computed with the {\sc hawk} program~\cite{hawk}, and
are small (3\% or less) for the Higgs boson mass range considered here.  The VBF cross sections in
Table~\ref{tab:higgsxsec} do not include the electroweak corrections.

In order to predict the distributions of the kinematics of Higgs boson signal events, CDF
and D0 use the \textsc{pythia}~\cite{pythia} Monte Carlo program,
with \textsc{CTEQ5L} and \textsc{CTEQ6L}~\cite{cteq} leading-order
(LO) parton distribution functions.

The Higgs boson decay branching fraction $Br(H\rightarrow\gamma\gamma)$ used here is
 obtained from the Handbook~\cite{lhcxs}.  The branching fractions
are computed by evaluating the partial decay widths of on-shell Higgs bosons to all possible final states
allowed in the SM, and evaluating the fractions of the total decay width for each process.  For all processes
except $H\rightarrow W^+W^-$ and $H\rightarrow ZZ$, {\sc hdecay}~\cite{hdecay}
 is used to compute the partial widths.  {\sc hdecay}
includes relevant higher-order QCD corrections to decays into quarks and gluons, and NLO electroweak corrections
are included in the $H\rightarrow \gamma\gamma$ and $H\rightarrow gg$ processes.  For $H\rightarrow W^+W^-$ and $H\rightarrow ZZ$,
the Monte Carlo generator {\sc Prophecy4f}~\cite{prophecy4f} is used.  It computes partial widths for Higgs boson decays at NLO,
including NLO QCD and electroweak corrections and all interferences, for four-fermion final states of Higgs boson decay.
More details are available in Ref.~\cite{lhcxs}.  The branching fraction $Br(H\rightarrow\gamma\gamma)$ is listed
in Table~\ref{tab:higgsxsec} as a function of $m_H$.


\begin{table}
\begin{center}
\caption{
The production cross sections and decay branching fractions
for the SM Higgs boson assumed for the combination.}
\vspace{0.2cm}
\label{tab:higgsxsec}
\begin{ruledtabular}
\begin{tabular}{cccccc}
$m_H$  & $\sigma_{gg\rightarrow H}$  & $\sigma_{WH}$  & $\sigma_{ZH}$  & $\sigma_{VBF}$    & $B(H\rightarrow \gamma\gamma)$ \\
(GeV/$c^2$) & (fb)                   & (fb)           & (fb)           & (fb)                & (\%) \\  \hline
   100 &   1821.8   &   291.90    &   169.8      &    100.1 & 0.159     \\
   105 &   1584.7   &   248.40    &   145.9      &     92.3 & 0.178     \\
   110 &   1385.0   &   212.00    &   125.7      &     85.1 & 0.197     \\
   115 &   1215.9   &   174.50    &   103.9      &     78.6 & 0.213     \\
   120 &   1072.3   &   150.10    &    90.2      &     72.7 & 0.225     \\
   125 &    949.3   &   129.50    &    78.5      &     67.1 & 0.230     \\
   130 &    842.9   &   112.00    &    68.5      &     62.1 & 0.226     \\
   135 &    750.8   &    97.20    &    60.0      &     57.5 & 0.214     \\
   140 &    670.6   &    84.60    &    52.7      &     53.2 & 0.194     \\
   145 &    600.6   &    73.70    &    46.3      &     49.4 & 0.168     \\
   150 &    539.1   &    64.40    &    40.8      &     45.8 & 0.137     \\
\end{tabular}	
\end{ruledtabular}	
\end{center}	
\end{table}

\section{Combining Channels} 
\label{sec:combination}

To gain confidence that the final result does not depend on the
details of the statistical formulation, we perform two types of
combinations, using Bayesian and Modified Frequentist approaches,
which yield limits on the Higgs boson production rate that agree
within 5\% at each value of $m_H$, and within 1\% on average.  Both
methods rely on distributions in the final discriminants, and not just
on their single integrated values.  Systematic uncertainties enter on
the predicted number of signal and background events as well as on the
distribution of the discriminants in each analysis (``shape
uncertainties'').  Both methods use likelihood calculations based on
Poisson probabilities.

Both methods treat the systematic uncertainties in a Bayesian fashion,
assigning a prior distribution to each source of uncertainty,
parameterized by a nuisance parameter, and propagating the impacts of
varying each nuisance parameter to the signal and background
predictions, with all correlations included.  A single nuisance
parameter may affect the signal and background predictions in many
bins of many different analyses.  Independent nuisance parameters are
allowed to vary separately within their prior distributions.
Both methods use the data to constrain the values of the nuisance
parameters, one by integration, the other by fitting.  These methods
reduce the impact of prior uncertainty in the nuisance parameters thus
improving the sensitivity.  Because of these constraints to the data,
it is important to evaluate the uncertainties and correlations
properly, and to allow independent parameters to vary separately,
otherwise a fit may overconstrain a parameter and extrapolate its use
improperly.  The impacts of correlated uncertainties add together
linearly on a particular prediction, while those of uncorrelated
uncertainties are convoluted together, which is similar to adding in
quadrature.  

\subsection{Bayesian Method}

Because there is no experimental information on the production cross
section for the Higgs boson, in the Bayesian
technique~\cite{prevhiggs}\cite{pdgstats} we assign a flat prior for
the total number of selected Higgs events.  For a given Higgs boson
mass, the combined likelihood is a product of likelihoods for the
individual channels, each of which is a product over histogram bins:

\begin{equation}
{\cal{L}}(R,{\vec{s}},{\vec{b}}|{\vec{n}},{\vec{\theta}})\times\pi({\vec{\theta}})
= \prod_{i=1}^{N_C}\prod_{j=1}^{N_b} \mu_{ij}^{n_{ij}} e^{-\mu_{ij}}/n_{ij}!
\times\prod_{k=1}^{n_{\rm{np}}}e^{-\theta_k^2/2}
\end{equation}

\noindent where the first product is over the number of channels
($N_C$), and the second product is over $N_b$ histogram bins containing
$n_{ij}$ events, binned in  ranges of the final discriminants used for
individual analyses, such as the reconstructed dijet mass or neural-network outputs.
 The parameters that contribute to the
expected bin contents are $\mu_{ij} =R \times s_{ij}({\vec{\theta}}) + b_{ij}({\vec{\theta}})$
for the
channel $i$ and the histogram bin $j$, where $s_{ij}$ and $b_{ij}$
represent the expected signal and background in the bin, and $R$ is a scaling factor
applied to the signal to test the sensitivity level of the experiment.
Truncated Gaussian priors are used for each of the $n_{\rm{np}}$ nuisance parameters
$\theta_k$, which define the sensitivity of the predicted signal and
background estimates to systematic uncertainties.  These can take the
form of uncertainties on overall rates, as well as the shapes of the
distributions used for combination.  These systematic uncertainties
can be far larger than the expected SM Higgs boson signal, and are
therefore important in the calculation of limits.  The truncation is
applied so that no prediction of any signal or background in any bin
is negative.  The posterior density function is then integrated over
all parameters (including correlations) except for $R$, and a 95\%
credibility level upper limit on $R$ is estimated by calculating the
value of $R$ that corresponds to 95\% of the area of the resulting
distribution.

\subsection{Modified Frequentist Method}

The Modified Frequentist technique relies on the ${\rm CL}_{\rm s}$
method, using a log-likelihood ratio (LLR) as test
statistic~\cite{DZHiggs}:
\begin{equation}
LLR = -2\ln\frac{p({\mathrm{data}}|H_1)}{p({\mathrm{data}}|H_0)},
\end{equation}
where $H_1$ denotes the test hypothesis, which admits the presence of
SM backgrounds and a Higgs boson signal, while $H_0$ is the null
hypothesis, for only SM backgrounds.  The probabilities $p$ are
computed using the best-fit values of the nuisance parameters for each
pseudo-experiment, separately for each of the two hypotheses, and
include the Poisson probabilities of observing the data multiplied by
Gaussian priors for the values of the nuisance parameters.  This
technique extends the LEP procedure~\cite{pdgstats} which does not
involve a fit, in order to yield better sensitivity when expected
signals are small and systematic uncertainties on backgrounds are
large~\cite{pflh}.

The ${\rm CL}_{\rm s}$ technique involves computing two $p$-values,
${\rm CL}_{\rm s+b}$ and ${\rm CL}_{\rm b}$.  The latter is defined by
\begin{equation}
1-{\rm CL}_{\rm b} = p(LLR\le LLR_{\mathrm{obs}} | H_0),
\end{equation}
where $LLR_{\mathrm{obs}}$ is the value of the test statistic computed
for the data. $1-{\rm CL}_{\rm b}$ is the probability of observing a
signal-plus-background-like outcome without the presence of signal,
i.e. the probability that an upward fluctuation of the background
provides a signal-plus-background-like response as observed in data.
The other $p$-value is defined by
\begin{equation}
{\rm CL}_{\rm s+b} = p(LLR\ge LLR_{\mathrm{obs}} | H_1),
\end{equation}
and this corresponds to the probability of a downward fluctuation of the sum
of signal and background in
the data.  A small value of ${\rm CL}_{\rm s+b}$ reflects inconsistency with  $H_1$.
It is also possible to have a downward fluctuation in data even in the absence of
any signal, and a small value of ${\rm CL}_{\rm s+b}$ is possible even if the expected signal is
so small that it cannot be tested with the experiment.  To minimize the possibility
of  excluding  a signal to which there is insufficient sensitivity
(an outcome  expected 5\% of the time at the 95\% C.L., for full coverage),
we use the quantity ${\rm CL}_{\rm s}={\rm CL}_{\rm s+b}/{\rm CL}_{\rm b}$.  If ${\rm CL}_{\rm s}<0.05$ for a particular choice
of $H_1$, that hypothesis is deemed to be excluded at the 95\% C.L. In an analogous
way, the expected ${\rm CL}_{\rm b}$, ${\rm CL}_{\rm s+b}$ and ${\rm CL}_{\rm s}$ values are computed from the median of the
LLR distribution for the background-only hypothesis.

Systematic uncertainties are included  by fluctuating the predictions for
signal and background rates in each bin of each histogram in a correlated way when
generating the pseudo-experiments used to compute ${\rm CL}_{\rm s+b}$ and ${\rm CL}_{\rm b}$.

\subsection{Systematic Uncertainties} 
\label{systematics}

Systematic uncertainties differ
between experiments and analyses, and affect the rates and shapes of the predicted
signal and background in correlated ways.  The combined results incorporate
the sensitivity of predictions to  values of nuisance parameters,
and include correlations between rates and shapes, between signals and backgrounds,
and between channels within experiments and between experiments.
More discussion on this topic can be found in the
individual analysis notes~\cite{cdfHgg} and~\cite{dzHgg}.  Here we
describe only the largest contributions and correlations between and
within the two experiments.

\subsubsection{Correlated Systematics between CDF and D0}

The uncertainties on the measurements of the integrated luminosities are 6\%
(CDF) and 6.1\% (D0).
Of these values, 4\% arises from the uncertainty
on the inelastic \pp~scattering cross section, which is correlated
between CDF and D0.

In both CDF and D0's $H\rightarrow\gamma\gamma$ analyses, the dominant background yields are calibrated with data control
samples, and thus the systematic uncertainties on the background rates and shapes are considered uncorrelated
between the two collaborations' results.

The theoretical cross section uncertainties on the inclusive $gg\rightarrow H$
production cross section ($\sim 14\%$), the WH and ZH cross sections ($\sim 6$\%),
and the VBF cross section ($\sim 5\%$) are shared between CDF and D0 and are used
in the computation of the limits scaled to the SM prediction, but are not
used in the calculation of the limit on $\sigma(p{\bar{p}}\rightarrow H)\times
Br(H\rightarrow\gamma\gamma)$.

\subsubsection{Correlated Systematic Uncertainties for CDF}
The dominant systematic uncertainties for the CDF analysis are shown
in Table~\ref{tab:cdfsystgg}.
Each source induces a
correlated uncertainty across all CDF channels' signal and background
contributions which are sensitive to that source.

\subsubsection{Correlated Systematic Uncertainties for D0 }
The dominant systematic uncertainties for the D0 analysis are shown in
Table~\ref{tab:d0systgg}.

\begin{table}[h]
\begin{center}
\caption{\label{tab:cdfsystgg} Systematic uncertainties on the signal contributions for CDF's
$H\rightarrow \gamma \gamma$ channels.
Systematic uncertainties are listed by name; see the original references
for a detailed explanation of their meaning and on how they are derived.  Uncertainties are relative, in
percent, and are symmetric unless otherwise indicated.}
\vskip 0.1cm
{\centerline{CDF: $H \rightarrow \gamma \gamma$ channel relative uncertainties (\%)}}
\vskip 0.099cm
\begin{ruledtabular}
\begin{tabular}{lcccc} \\
Channel & CC & CP & CC Conv & CP Conv \\ \hline
 & \multicolumn{4}{c}{Systematic Uncertainties on Signal (\%)} \\
Luminosity & 6 & 6 & 6 & 6 \\
$\sigma_{ggH}/\sigma_{VH}/\sigma_{VBF}$ &
    14/7/5 & 14/7/5 & 14/7/5 & 14/7/5 \\
 PDF & 2 & 2 & 2 & 2 \\
 ISR & 3 & 4 & 2 & 5 \\
 FSR & 3 & 4 & 2 & 5 \\
Energy Scale & 0.2 & 0.8 & 0.1 & 0.8 \\
Trigger Efficiency & -- & -- & 0.1 & 0.4 \\
$z$ Vertex & 0.2 & 0.2 & 0.2 & 0.2 \\
Conversion ID & -- & -- & 7 & 7 \\
Detector Material & 0.4 & 3.0 & 0.2 & 3.0 \\
Photon/Electron ID & 1.0 & 2.8 & 1.0 & 2.6 \\
Run Dependence & 3.0 & 2.5 & 1.5 & 2.0 \\
Data/MC Fits & 0.4 & 0.8 & 1.5 & 2.0 \\ \hline
 & \multicolumn{4}{c}{Systematic Uncertainties on Background (\%)} \\
Fit Function & 3.5 & 1.1 & 7.5 & 3.5 \\
\end{tabular}
\end{ruledtabular}
\end{center}
\end{table}

\begin{table}[h]
\begin{center}
\caption{\label{tab:d0systgg} Systematic uncertainties on the signal and background contributions for D0's
$H\rightarrow \gamma \gamma$ channel. Systematic uncertainties for the Higgs signal shown in this table are
obtained for $m_H=125$ GeV/$c^2$.  Systematic uncertainties are listed by name; see the original references
for a detailed explanation of their meaning and on how they are derived.  Uncertainties are relative, in
percent, and are symmetric unless otherwise indicated.}
\vskip 0.1cm
{\centerline{D0: $H \rightarrow \gamma \gamma$ channel relative uncertainties (\%)}}
\vskip 0.099cm
\begin{ruledtabular}
\begin{tabular}{lcc}\\
Contribution &  ~~~Background~~~  & ~~~Signal~~~    \\
\hline
Luminosity~~~~                            &  6     &  6    \\
Acceptance                                &  --    &  2    \\
electron ID efficiency                    &  2     &  --   \\
electron track-match inefficiency         & 10     &  --   \\
Photon ID efficiency                      &  3     &   3   \\
Photon energy scale                       &  2     &   1   \\
Cross Section                             &  4     &  10   \\
Background subtraction                    &  8 &  -       \\
\end{tabular}
\end{ruledtabular}
\end{center}
\end{table}

\section{Combined Results} 

Using the combination procedures outlined in Section III, we extract
limits on SM Higgs boson production $\sigma \times B(H\rightarrow\gamma\gamma)$
in \pp~collisions at $\sqrt{s}=1.96$~TeV for $100\leq m_H \leq 150$
GeV/$c^2$.  To facilitate comparisons with the SM and to
accommodate analyses with different degrees of sensitivity, we present
our results in terms of the ratio of obtained limits to the SM Higgs
boson production cross section, as a function of Higgs boson mass, for
test masses arranged in 5~GeV/$c^2$ steps between 100 and 150~GeV/$c^2$.

The ratios of the 95\% C.L. expected and observed limit to the SM
cross section are shown in Figure~\ref{fig:comboRatio} for the
combined CDF and D0 analyses.  The observed and median expected ratios
computed using the Bayesian and the ${\rm CL}_{\rm s}$ methods
are listed for the tested Higgs boson masses in Table~\ref{tab:ratios}.
In the following summary we quote only the limits obtained
with the Bayesian method, which was chosen {\it a priori}.
We obtain the observed (expected) values of 7.2~(10.6) at
$m_{H}=100$~GeV/$c^2$, 10.5~(8.5) at $m_{H}=115$~GeV/$c^2$, and
13.4~(8.3) at $m_{H}=120$~GeV/$c^2$.

We also compute limits on the inclusive production cross section times
the decay branching ratio $\sigma(p{\bar{p}}\rightarrow H+X)\times Br(H\rightarrow\gamma\gamma)$,
in order to reduce model dependence from the SM predictions of the cross section and the decay
branching ratio.  Some residual model dependence remains, however, as the SM ratios of
$WH, ZH,$ VBF, and $gg\rightarrow H$ production are assumed in order to compute the
signal acceptances for each channel.  The signal acceptances vary by no more than a relative 10\% between
the different production mechanisms, however, and so the sensitivity to the predicted ratios is small.
We perform the combination calculation for this result
without the theoretical uncertainties on the total production cross section and the decay branching
fraction $Br(H\rightarrow\gamma\gamma)$. 
The resulting limits are shown in Figure~\ref{fig:xsbrlimits} and listed in Table~\ref{tab:xsbrlimits}.

\begin{figure}[hb]
\begin{centering}
\includegraphics[width=16.5cm]{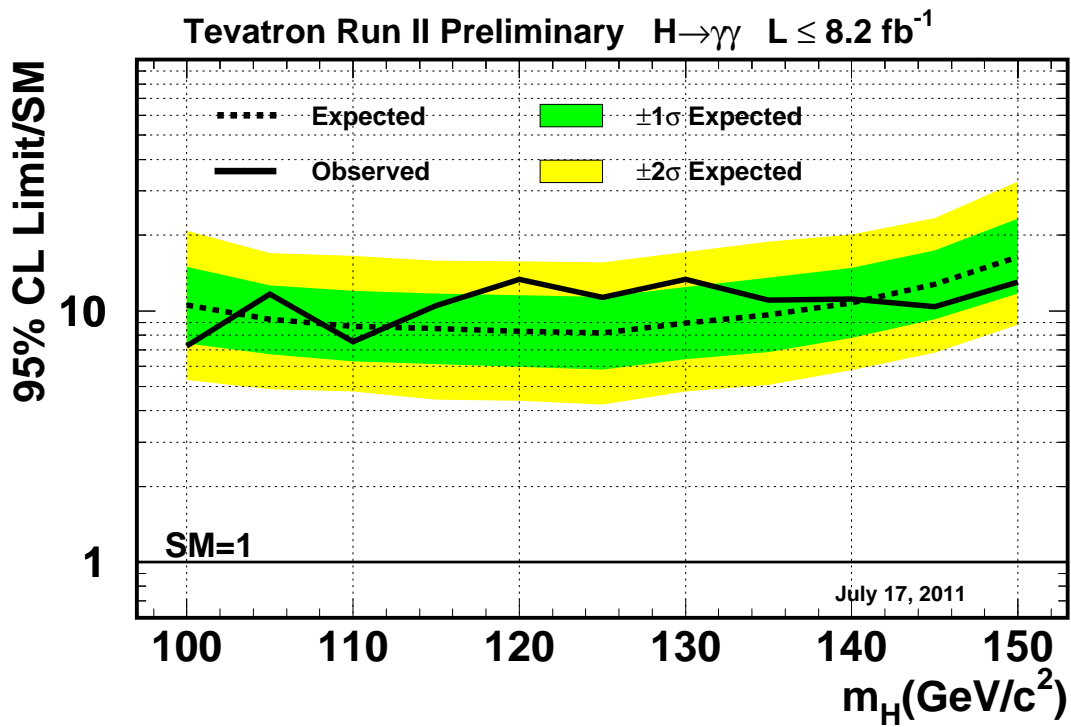}
\caption{
\label{fig:comboRatio}
Observed and expected (median, for the background-only hypothesis)
95\% C.L. upper limits on the ratios to the SM cross section, as
functions of the Higgs boson mass for the combined CDF and D0
analyses.  The limits are expressed as a multiple of the SM prediction
for test masses (every 5 GeV/$c^2$) for which both experiments have
performed dedicated searches in different channels.  The points are
joined by straight lines for better readability.  The bands indicate
the 68\% and 95\% probability regions where the limits can fluctuate,
in the absence of signal.  The limits displayed in this figure are
obtained with the Bayesian calculation.  }
\end{centering}
\end{figure}

\begin{table}[ht]
\caption{\label{tab:ratios} Ratios of median expected and observed 95\% C.L.
limit to the SM cross section prediction for the combined CDF and D0 analyses as a function
of the Higgs boson mass in GeV/$c^2$, obtained with the Bayesian and  ${\rm CL}_{\rm s}$ methods.}
\begin{ruledtabular}
\begin{tabular}{lccccccccccc}\\
Bayesian                   &  100 &  105 &  110 &  115  &  120 & 125 & 130 & 135 & 140  & 145  & 150\\ \hline
Expected                   & 10.6 & 9.2 &  8.7 &  8.5  &  8.3 & 8.2 & 8.9 &9.6 & 10.8 & 12.7 & 16.4 \\
Observed                   & 7.2  & 11.7 &  7.6 &  10.5 &  13.4& 11.3& 13.3 &11.0 & 11.1 & 10.4 & 13.0 \\
\hline
\hline\\
${\rm CL}_{\rm s}$         &  100 &  105 &  110 &  115  &  120 & 125 & 130 & 135 & 140  & 145  & 150\\ \hline
Expected:                  &10.5& 9.3 & 8.9 & 8.5 & 8.2 & 8.2 & 8.7 & 9.4 & 10.7& 12.9& 16.7 \\
Observed:                  &7.0   &11.7  &7.6   &10.5  &13.3  &11.3 & 13.5&11.1 & 11.2 & 10.2 & 12.7  \\
\end{tabular}
\end{ruledtabular}
\end{table}

\begin{table}[ht]
\caption{\label{tab:xsbrlimits}
The 95\% CL limit on the product of the inclusive Higgs boson
production cross section and the decay branching ratio to a pair of photons
$\sigma(p{\bar{p}}\rightarrow H+X)\times Br(H\rightarrow\gamma\gamma)$, as a function
of the Higgs boson mass, assuming SM ratios of the four production mechanisms.}
\begin{ruledtabular}
\begin{tabular}{lccccccccccc}\\
Bayesian                   &  100 &  105 &  110 &  115  &  120 & 125 & 130 & 135 & 140  & 145  & 150\\ \hline
Expected (fb)              & 40.1 & 33.7 & 31.7 &  28.2 & 25.5 & 23.2& 21.5& 19.5& 18.2 & 16.4 & 15.6 \\
Observed (fb)              & 27.2 & 42.6 & 26.8 &  34.9 & 42.0 & 31.9& 32.6& 22.5& 18.6 & 13.6 & 12.4 \\
\hline
\end{tabular}
\end{ruledtabular}
\end{table}

\begin{figure}[hb]
\begin{centering}
\includegraphics[width=16.5cm]{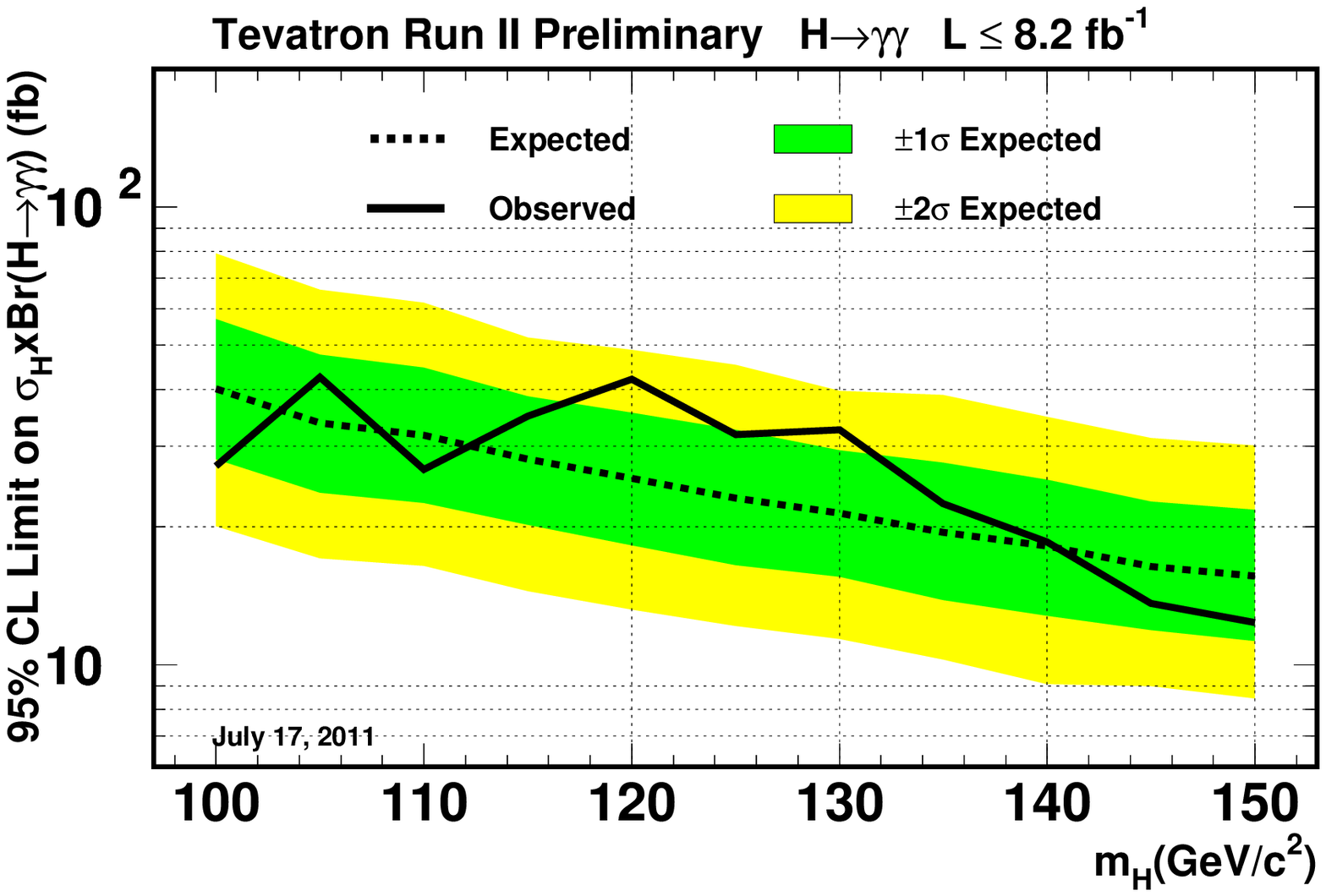}
\caption{
\label{fig:xsbrlimits}
Limits on the product of the inclusive Higgs boson
production cross section and the decay branching ratio to a pair of photons
$\sigma(p{\bar{p}}\rightarrow H+X)\times Br(H\rightarrow\gamma\gamma)$, as a function
of the Higgs boson mass, assuming SM ratios of the four production mechanisms. The bands indicate
the 68\% and 95\% probability regions where the limits can fluctuate,
in the absence of signal.}
\end{centering}
\end{figure}

In summary, we combine CDF and D0 results on SM Higgs boson searches with $H\rightarrow\gamma\gamma$,
based on 7.0 fb$^{-1}$ of data from CDF and 8.2 fb$^{-1}$ of data from D0.
We use the recommendation of the PDF4LHC
working group for the central value of the parton distribution functions and uncertainties~\cite{pdf4lhc}.
We use the highest-order calculations available for the $gg \rightarrow H$, $WH$, $ZH$, and VBF theoretical cross
sections when comparing our limits to the SM predictions.  We include consensus estimates
of the theoretical uncertainties on these production cross sections and the decay
branching fractions in the computations of our limits.

The 95\% C.L. upper limit on Higgs boson production is a factor of 10.5
times the SM cross section for a Higgs boson mass of $m_{H}=$115~GeV/$c^2$.
Based on simulation, the corresponding median expected upper limit is 8.5 times the SM cross section.
Standard Model branching ratios, calculated as functions of the Higgs boson mass, are assumed.
The results presented here extend significantly the sensitivity of the separate CDF and D0 results.

\begin{center}
{\bf Acknowledgments}
\end{center}

We thank the Fermilab staff and the technical staffs of the
participating institutions for their vital contributions, and we
acknowledge support from the
DOE and NSF (USA);
CONICET and UBACyT (Argentina);
ARC (Australia);
CNPq, FAPERJ, FAPESP and FUNDUNESP (Brazil);
CRC Program and NSERC (Canada);
CAS, CNSF, and NSC (China);
Colciencias (Colombia);
MSMT and GACR (Czech Republic);
Academy of Finland (Finland);
CEA and CNRS/IN2P3 (France);
BMBF and DFG (Germany);
INFN (Italy);
DAE and DST (India);
SFI (Ireland);
Ministry of Education, Culture, Sports, Science and Technology (Japan);
NRF, KRF, KOSEF, and the World Class University Program (Korea);
CONACyT (Mexico);
FOM (The Netherlands);
FASI, Rosatom and RFBR (Russia);
Slovak R\&D Agency (Slovakia); 
Ministerio de Ciencia e Innovaci\'{o}n, and Programa Consolider-Ingenio 2010 (Spain);
The Swedish Research Council (Sweden);
Swiss National Science Foundation (Switzerland);
STFC and the Royal Society (United Kingdom);
and
the A.P. Sloan Foundation (USA).



\clearpage
\begin{thebibliography}{000}

\bibitem{atlasww} ATLAS Collaboration, ``Higgs Boson Searches using the
$H\rightarrow WW^{(*)}\rightarrow\ell\nu\ell\nu$ Decay Mode with the ATLAS Detector
at 7~TeV'', ATLAS-CONF-2011-005 (2011).

\bibitem{cmsww} CMS Collaboration, 
  ``Measurement of W+W- Production and Search for the Higgs Boson in pp Collisions at sqrt(s) = 7 TeV,''
    [arXiv:1102.5429 [hep-ex]] (2011).

\bibitem{atlaszz} ATLAS Collaboration, ``Search for a Standard Model Higgs Boson in the Mass Range
200-600 GeV in the Channels
$H\rightarrow ZZ\rightarrow\ell^+\ell^-\nu{\bar{\nu}}$ and
$H\rightarrow ZZ\rightarrow\ell^+\ell^-q{\bar{q}}$ with the ATLAS Detector'',
ATLAS-CONF-2011-026 (2011).

\bibitem{atlasgammagamma} ATLAS Collaboration, ``Search for the Higgs boson in the diphoton
final state with 37.6 pb$^{-1}$ of data recorded by the ATLAS detector
in proton-proton collisions at $\sqrt{s}=7$~TeV'', ATLAS-CONF-2011-025 (2011).


\bibitem{atlasgammagammaupdate} ATLAS Collaboration, ``Update of the Background Studies in the Search for the Higgs Boson
in the Two Photons Channel in $pp$ Collisions at $\sqrt{s}$=7 TeV'',
ATLAS-CONF-2011-071 (2011).


\bibitem{cdfHWW} CDF Collaboration, ``Search for $H \rightarrow WW^*$ Production Using 7.1~fb$^{-1}$'',
CDF Conference Note 10432 (2011).

\bibitem{DZHiggs} \DZero Collaboration, ``Combined Upper Limits on Standard Model Higgs Boson Production in the $W^{+}W^{-}$, $\tau\tau$ and $\gamma\gamma$ decay modes from the D0 Experiment in up to 8.2 fb$^{-1}$ of data'',
D0 Conference Note 6183 (2011).

\bibitem{prevcomb} The CDF and D0 Collaborations and the TEVNPHWG Working Group,
``Combined CDF and D0 Upper Limits on Standard Model Higgs-Boson Production with up to 8.2 fb$^{-1}$ of Data'',
FERMILAB-CONF-11-044-E, CDF Note 10441, D0 Note 6184, arXiv:1103.3233v2 [hep-ex] (2011).

\bibitem{prevhiggs}
The CDF and D0 Collaborations and the TEVNPHWG Working Group,
``Combined CDF and D0 Upper Limits on Standard Model Higgs-Boson Production with up to 6.7 fb$^{-1}$ of Data'',
FERMILAB-CONF-10-257-E, CDF Note 10241, D0 Note 6096, arXiv:1007.4587v1 [hep-ex] (2010);  \\
CDF Collaboration, ``Combined Upper Limit on Standard Model Higgs Boson 
Production for ICHEP 2010'', CDF Conference Note 10241 (2010); \\
\DZero Collaboration, ``Combined Upper Limits on Standard Model Higgs Boson Production from the D0 Experiment
in up to 6.7 fb$^{-1}$ of data'', D0 Conference Note 6094 (2010).

The CDF and D0 Collaborations and the TEVNPHWG Working Group,
``Combined CDF and DZero Upper Limits on Standard Model Higgs-Boson Production with 2.1 to 5.4 fb$^{-1}$ of Data'',
FERMILAB-PUB-09-0557-E, CDF Note 9998, D0 Note 5983,
arXiv:0911.3930v1 [hep-ex] (2009); \\
CDF Collaboration, ``Combined Upper Limit on Standard Model Higgs Boson Production for HCP 2009'', CDF Conference Note 9999 (2009); \\
\DZero Collaboration,``Combined Upper Limits on Standard Model Higgs Boson Production from the D0 Experiment
in 2.1-5.4 fb$^{-1}$'', D0 Conference Note 6008 (2009).


\bibitem{WWPRLhiggs} CDF Collaboration, ``Inclusive Search for Standard Model Higgs Boson Production in the
WW Decay Channel Using the CDF II Detector'', Phys. Rev. Lett. 104, 061803 (2010); \\
\DZero Collaboration, `` Search for Higgs Boson Production in Dilepton and Missing Energy Final States with 5.4
fb$^{-1}$ of $p\bar{p}$ Collisions at $\sqrt{s}=1.96$~TeV'', Phys. Rev. Lett. 104, 061804 (2010); \\
The CDF and D0 Collaborations, ``Combination of Tevatron Searches for the Standard Model Higgs Boson in the
$W^+W^-$ Decay Mode'', Phys. Rev. Lett. 104, 061802 (2010).

\bibitem{cdfHgg} CDF Collaboration, ``Search for a Standard Model Higgs Boson Decaying Into
Photons at CDF Using 7.0~fb$^{-1}$ of Data'', CDF Note 10485 (2011).

\bibitem{dzHgg} D0 Collaboration, ``Search for the standard model and a fermiophobic Higgs boson in diphoton final states'',
[e-Print: arXiv:1107.4587 [hep-ph]].

\bibitem{anastasiou}
 C.~Anastasiou, R.~Boughezal and F.~Petriello,
  JHEP {\bf 0904}, 003 (2009).

\bibitem{grazzinideflorian}
 D.~de Florian and M.~Grazzini,
  Phys.\ Lett.\  B {\bf 674}, 291 (2009).

\bibitem{grazziniprivate} M.~Grazzini, private communication (2010).

\bibitem{tevtop10}  The CDF and D0 Collaborations and the Tevatron Electroweak Working Group,
arXiv:1007.3178 [hep-ex], arXiv:0903.2503~[hep-ex].

\bibitem{harlanderkilgore2002} R.~V.~Harlander and W.~B.~Kilgore, Phys. Rev. Lett. {\bf 88}, 201801 (2002).

\bibitem{anastasioumelnikov2002} C.~Anastasiou and K.~Melnikov, Nucl. Phys. B~{\bf 646}, 220 (2002).

\bibitem{ravindran2003} V.~Ravindran, J.~Smith, and W.~L. van~Neerven, Nucl. Phys. B~{\bf 665}, 325 (2003).

\bibitem{actis2008} S.~Actis, G.~Passarino, C.~Sturm, and S.~Uccirati, Phys. Lett. B~{\bf 670}, 12 (2008).

\bibitem{aglietti} U. Aglietti, R. Bonciani, G. Degrassi, A. Vicini, ``Two-loop electroweak corrections to Higgs
production in proton-proton collisions'', arXiv:hep-ph/0610033v1 (2006).

\bibitem{catani2003}
S.~Catani, D.~de Florian, M.~Grazzini and P.~Nason,
   ``Soft-gluon resummation for Higgs boson production at hadron colliders,''
  JHEP {\bf 0307}, 028 (2003)
  [arXiv:hep-ph/0306211].

\bibitem{mstw2008}
 A.~D.~Martin, W.~J.~Stirling, R.~S.~Thorne and G.~Watt,
  Eur.\ Phys.\ J.\  C {\bf 63}, 189 (2009).


\bibitem{pdf4lhc} {\url{http://www.hep.ucl.ac.uk/pdf4lhc/}}; \\
S. Alekhin {\it et al}., (PDF4LHC Working Group), [arXiv:1101.0536v1 [hep-ph]]; \\
M. Botje {\it et al}., (PDF4LHC Working Group), [arXiv:1101.0538v1 [hep-ph]].

\bibitem{cteq66}    P.~M.~Nadolsky {\it et al.},
  Phys.\ Rev.\  D {\bf 78}, 013004 (2008)
  [arXiv:0802.0007 [hep-ph]].

\bibitem{nnpdf}  R.~D.~Ball {\it et al.}  [NNPDF Collaboration],
  Nucl.\ Phys.\  B {\bf 809}, 1 (2009)
  [Erratum-ibid.\  B {\bf 816}, 293 (2009)]
  [arXiv:0808.1231 [hep-ph]].

\bibitem{djouadibaglio}  J.~Baglio and A.~Djouadi,
  JHEP {\bf 1010}, 064 (2010)
  [arXiv:1003.4266v2 [hep-ph]].


\bibitem{v2hv} The Fortran program can be found on Michael Spira's web page
{\tt http://people.web.psi.ch/$\sim$mspira/proglist.html}.

\bibitem{vhnnloqcd} O.~Brein, A.~Djouadi, and R.~Harlander, Phys. Lett. B {\bf 579}, 149 (2004).

\bibitem{vhewcorr} M.~L.~Ciccolini,  S.~Dittmaier, and M.~Kramer, Phys. Rev. D {\bf 68}, 073003 (2003).

\bibitem{vbfnnlo} P.~Bolzoni, F.~Maltoni, S.-O.~Moch, and M.~Zaro,
Phys. Rev. Lett. {\bf 105}, 011801 (2010)
[arXiv:1003.4451v2 [hep-ph]].

\bibitem{hawk} M.~Ciccolini, A.~Denner, and S.~Dittmaier,
Phys. Rev. Lett. {\bf 99}, 161803 (2007)
[arXiv:0707.0381 [hep-ph]]; \\
M.~Ciccolini, A.~Denner, and S.~Dittmaier,
Phys. Rev. D {\bf 77}, 013002 (2008) [arXiv:0710.4749 [hep-ph]].\\
We would like to thank the authors of the {\sc hawk} program for adapting it to the Tevatron.

\bibitem{pythia}
T.~Sj\"ostrand, L.~Lonnblad and S.~Mrenna,
   ``PYTHIA 6.2: Physics and manual,''
  arXiv:hep-ph/0108264.

\bibitem{cteq}
H.~L.~Lai {\it et al.},
Phys. Rev D \textbf{55}, 1280 (1997).


\bibitem{lhcxs} S.~Dittmaier~{\it et al.} [LHC Higgs Cross Section Working Group],
``Handbook of LHC Higgs Cross Sections: 1. Inclusive Observables'',
arXiv:1101.0593v2 (2011).

\bibitem{hdecay}
A.~Djouadi, J.~Kalinowski and M.~Spira,
  Comput.\ Phys.\ Commun.\  {\bf 108}, 56 (1998).

\bibitem{prophecy4f} A.~Bredenstein, A.~Denner, S.~Dittmaier, and M.~M.~Weber,
Phys. Rev. D {\bf 74}, 013004 (2006); \\
A. Bredenstein, A.~Denner, S.~Dittmaier, and M.~Weber, JHEP {\bf 0702}, 080 (2007); \\
A. Bredenstein, A.~Denner, S.~Dittmaier, A. M\"uck, and M.~M.~Weber, JHEP {\bf 0702}, 080 (2007)
{\url{http://omnibus.uni-freiburg.de/~sd565/programs/prophecy4f/prophecy4f.html}} (2010).


\bibitem{pdgstats}
T. Junk, Nucl. Instrum. Meth. A {\bf 434}, 435 (1999); \\
A.L.~Read, ``Modified Frequentist analysis of search results (the ${\rm CL}_{\rm s}$ method)'', in
F.~James, L.~Lyons and Y.~Perrin (eds.), {\sl Workshop on Confidence Limits},
CERN, Yellow Report 2000-005, available through {\tt cdsweb.cern.ch}.

\bibitem{pflh} W.~Fisher, ``Systematics and Limit Calculations,''
FERMILAB-TM-2386-E.


\end{thebibliography}
\end{document}